\documentstyle[12pt]{article}
\def\emline#1#2#3#4#5#6{%
       \put(#1,#2){\special{em:moveto}}%
       \put(#4,#5){\special{em:lineto}}}
\def\newpic#1{}

\title{Multidimensional analogs of geometric \goodbreak
$s\leftrightarrow t$ duality}
\author{I.G. Korepanov\thanks{South Ural State University,
76 Lenin avenue, Chelyabinsk 454080, Russia.
 E-Mail: igor@prima.tu-chel.ac.ru}}
\date{}

\def\p{\partial}

\def\be{\begin{equation}}
\def\ee{\end{equation}}

\begin{document}
\maketitle

\begin{abstract}
The usual propetry of $s\leftrightarrow t$ duality for scattering amplitudes,
e.g.\ for Veneziano amplitude, is deeply connected with the 2-dimensional
geometry. In particular, a simple geometric construction of such amplitudes
was proposed in a joint work by this author and S.~Saito (solv-int/9812016).
Here we propose analogs of one of those amplitudes associated
with multidimensional euclidean spaces, paying most attention
to the 3-dimensional case. Our results can be regarded as a variant of
``Regge calculus'' intimately connected with ideas of the theory
of integrable models.
\end{abstract}

\section*{Introduction}

Mathematical physics was very successful in investigating two-dimensional
(or ``$1+1$-dimensional") systems of most various kinds. The next step
must be as deep investigation of 3-dimensional, and then multidimensional
systems. The transfer of interest to 3-dimensional systems is observed
particularly well in the theory of exactly solvable models, where the
research is now focussed on the {\em tetrahedron equation} ---
the 3-dimensional analog of the famous Yang -- Baxter equation.

A remarkable achievement of that theory was demonstration of the fact that
virtually all usual classical solitonic systems arise from a simple
wye -- delta transformation in electric circuits known since
1899~\cite{kennelly}. This was shown in paper~\cite{kashaev}, and
in~\cite{KKS} one can find a matrix generalization of that construction,
including the quantum case, based on the ideas of earlier
papers~\cite{zap.pomi,dokt.dis}.

A domain close to exactly solvable models is studying of scattering
amplitudes that satisfy the $s\leftrightarrow t$ duality. Remarkably,
here, too, some elementary geometric constructions were found recently
that produce such amplitudes~\cite{KS}. These constructions admit as well
wide generalizations related to (almost) arbitrary Lie groups.

Nevertheless, the constructions of~\cite{KS} appear naturally
associated with geometry of {\em 2-dimensional\/} spaces. Recall by the way
that it was from Veneziano amplitude~\cite{veneziano} obeying
$s\leftrightarrow t$ duality that the string theory started
with its 2-dimensional string world sheet. It is clear that search
for similar constructions associated with multidimensional spaces
in a natural and important problem.

In the present paper a generalization is proposed to one of the amplitudes
in~\cite{KS}, namely the amplitude associated with the 2-dimensional
euclidean space,
to the 3-dimensional case and then to the case of any dimension.
We replace triangles from~\cite{KS} associated with ``elementary"
amplitudes by tetrahedra or, respectively, higher dimensional simplices,
and a composition of several amplitudes is represented as
a simplicial complex. The result can be regarded as some version of
Regge calculus linked conceptually, due to a generalized $s\leftrightarrow t$
duality, to the theory of integrable models.

Below, in section~\ref{sec 2-dim} we re-derive the formula for
``2-dimensional" amplitude from~\cite{KS} in the manner that makes
the way to generalizations most clear. Then we compute a composition
of amplitudes and study duality transformations on the example
of a complex (hexagon) made of four triangles.
Section~\ref{sec 2-dim} is intended
to facilitate, by analogy, understanding of the main section~\ref{sec 3-dim}
where we are occupied with 3-dimensional and then multidimensional
constructions. Finally, section~\ref{questions} is devoted to a discussion,
mostly in the form of a list of questions for further work.

\section{Amplitudes related to the 2-dimensional euclidean space}
\label{sec 2-dim}

In this section, we will start with re-deriving one of the formulae
of~\cite{KS}. Our calculations, very elementary, will be performed
is a manner that makes most obvious the feasibility of generalizations.
Note, by the way, that the most obvious seems the possibility to change
the group of motions of euclidean plane to other groups, already sketched
in the last section of~\cite{KS}. More fascinating, however, looks
the ``multidimensional" generalization proposed below in
section~\ref{sec 3-dim}.

\subsection{Duality from a euclidean quadrangle}

Consider a quadrangle $ABCD$ in a 2-dimensional euclidean space
(fig.~\ref{fig 4-ugol'nik}).
\begin{figure}
\begin{center}
\unitlength=1mm
\special{em:linewidth 0.4pt}
\linethickness{0.4pt}
\begin{picture}(34.00,31.00)
\emline{23.00}{3.00}{1}{33.00}{18.00}{2}
\emline{33.00}{18.00}{3}{18.00}{30.00}{4}
\emline{18.00}{30.00}{5}{2.00}{18.00}{6}
\emline{2.00}{18.00}{7}{23.00}{3.00}{8}
\emline{23.00}{3.00}{9}{18.00}{30.00}{10}
\emline{3.00}{18.00}{11}{33.00}{18.00}{12}
\put(1.00,18.00){\makebox(0,0)[rc]{$A$}}
\put(18.00,31.00){\makebox(0,0)[cb]{$D$}}
\put(34.00,18.00){\makebox(0,0)[lc]{$C$}}
\put(23.00,2.00){\makebox(0,0)[ct]{$B$}}
\put(26.00,25.00){\makebox(0,0)[lb]{$l_5$}}
\put(9.00,25.00){\makebox(0,0)[rb]{$l_1$}}
\put(11.50,10.50){\makebox(0,0)[rt]{$l_2$}}
\put(28.00,9.00){\makebox(0,0)[lt]{$l_4$}}
\put(12.00,19.00){\makebox(0,0)[cb]{$l_6$}}
\put(22.00,12.00){\makebox(0,0)[lb]{$l_3$}}
\end{picture}
\end{center}
\caption{}
\label{fig 4-ugol'nik}
\end{figure}
Let the edge lengths $l_1,l_2,l_4$ and $l_5$ be fixed, while the diagonals
$l_3$ and $l_6$ be able to change. We want to compute the derivative
$\,d l_6/\,d l_3$.

We will do this by ``making a road from $l_3$ to $l_6$
via $l_1$". By these words, we mean the use of the implicit function
derivative formula
\be
{\,d l_6 \over \,d l_3}=-{\p l_1/\p l_3 \over \p l_1/\p l_6},
\label{eq cherez l1}
\ee
where $l_1$ is regarded as a fuction of all other edge lengths.

To consider the ``first half of the road", we will assume for a moment that
all the lengths {\em except $l_3$ and $l_1$} are fixed. Then
\be
{\p l_1 \over \p l_3}={\,d l_1 \over \,d l_3}=
{\,d l_1 \over \,d \angle ACD}\cdot {\,d \angle ACD \over \,d \angle BCD}
\cdot {\,d \angle BCD \over \,d \l_3}.
\label{eq dl1/dl3}
\ee

The second factor in the RHS of~(\ref{eq dl1/dl3}) equals unity,
while the first and the third --- $2S_{ACD}/l_1$ and $l_3/(2S_{BCD})$
respectively, where $S_{\ldots}$ denotes the area of a respective
triangle. From here we find
\be
{l_1 \,d l_1 \over S_{ACD}}={l_3 \,d l_3 \over S_{BCD}}.
\label{eq l3-l1}
\ee

Moving similarly further to $l_6$ and taking (\ref{eq cherez l1})
into account, we find
\be
{l_6 \,d l_6 \over S_{ABC}\cdot S_{ACD}}=
-{l_3 \,d l_3 \over S_{ABD}\cdot S_{BCD}}.
\label{eq l3-l6}
\ee
This is, essentially, what we call the duality relation. As explained
in~\cite{KS}, putting in correspondence to a triangle of area~$S$
and sides $k,l,m$ the amplitude
\be
A(k,l,m)=\exp(\lambda S)/S
\label{ampl s exp}
\ee
(the exponential factor can be introduced due to the relation
$$
S_{ABC}+S_{ACD}=S_{ABD}+S_{BCD},
$$
with the same constant $\lambda$ for all triangles), we get, with a
proper choice of integration limits, the duality in the form
\be
\intop A(k,l,s)\, A(s,m,n) \,d\mu(s)=
\intop A(k,m,t)\, A(t,l,n) \,d\mu(t),
\label{eq 5.5}
\ee
where
\be
\,d\mu(l)=l \,d l.
\label{d mu}
\ee

We propose to integrate in formula~(\ref{eq 5.5}) over
{\em all\/} possible configurations of a flat euclidean
quadrangle, including those nonconvex and with self-intersections,
taking as the integrand the {\em
absolute value\/} of LHS or RHS of (\ref{eq l3-l6}), i.e.\ assuming
that both the areas of triangles and the integration measures
(\ref{d mu}) are always positive (note that in~\cite{KS}
another possibility was considered).

\subsection{An example of a greater complex}

Consider an example depicted in fig.~\ref{fig 6-ugol'niki},
\begin{figure}
\begin{center}
\unitlength=1.00mm
\special{em:linewidth 0.4pt}
\linethickness{0.4pt}
\begin{picture}(95.00,16.00)
\emline{2.00}{5.00}{1}{8.00}{2.00}{2}
\emline{8.00}{2.00}{3}{14.00}{5.00}{4}
\emline{14.00}{5.00}{5}{14.00}{12.00}{6}
\emline{14.00}{12.00}{7}{8.00}{15.00}{8}
\emline{8.00}{15.00}{9}{2.00}{12.00}{10}
\emline{2.00}{12.00}{11}{2.00}{5.00}{12}
\emline{2.00}{5.00}{13}{14.00}{5.00}{14}
\emline{14.00}{5.00}{15}{8.00}{15.00}{16}
\emline{8.00}{15.00}{17}{2.00}{5.00}{18}
\emline{23.00}{5.00}{19}{29.00}{2.00}{20}
\emline{29.00}{2.00}{21}{35.00}{5.00}{22}
\emline{35.00}{5.00}{23}{35.00}{12.00}{24}
\emline{35.00}{12.00}{25}{29.00}{15.00}{26}
\emline{29.00}{15.00}{27}{23.00}{12.00}{28}
\emline{23.00}{12.00}{29}{23.00}{5.00}{30}
\emline{23.00}{5.00}{31}{35.00}{5.00}{32}
\emline{29.00}{15.00}{33}{23.00}{5.00}{34}
\emline{42.00}{5.00}{35}{48.00}{2.00}{36}
\emline{48.00}{2.00}{37}{54.00}{5.00}{38}
\emline{54.00}{5.00}{39}{54.00}{12.00}{40}
\emline{54.00}{12.00}{41}{48.00}{15.00}{42}
\emline{48.00}{15.00}{43}{42.00}{12.00}{44}
\emline{42.00}{12.00}{45}{42.00}{5.00}{46}
\emline{42.00}{5.00}{47}{54.00}{5.00}{48}
\emline{61.00}{5.00}{49}{67.00}{2.00}{50}
\emline{67.00}{2.00}{51}{73.00}{5.00}{52}
\emline{73.00}{5.00}{53}{73.00}{12.00}{54}
\emline{73.00}{12.00}{55}{67.00}{15.00}{56}
\emline{67.00}{15.00}{57}{61.00}{12.00}{58}
\emline{61.00}{12.00}{59}{61.00}{5.00}{60}
\emline{82.00}{5.00}{61}{88.00}{2.00}{62}
\emline{88.00}{2.00}{63}{94.00}{5.00}{64}
\emline{94.00}{5.00}{65}{94.00}{12.00}{66}
\emline{94.00}{12.00}{67}{88.00}{15.00}{68}
\emline{88.00}{15.00}{69}{82.00}{12.00}{70}
\emline{82.00}{12.00}{71}{82.00}{5.00}{72}
\emline{23.00}{5.00}{73}{35.00}{12.00}{74}
\emline{42.00}{12.00}{75}{54.00}{12.00}{76}
\emline{54.00}{12.00}{77}{42.00}{5.00}{78}
\emline{61.00}{12.00}{79}{73.00}{12.00}{80}
\emline{73.00}{12.00}{81}{61.00}{5.00}{82}
\emline{73.00}{12.00}{83}{67.00}{2.00}{84}
\emline{82.00}{12.00}{85}{94.00}{12.00}{86}
\emline{94.00}{12.00}{87}{88.00}{2.00}{88}
\emline{88.00}{2.00}{89}{82.00}{12.00}{90}
\put(18.00,8.50){\vector(1,0){3.00}}
\put(37.00,8.50){\vector(1,0){3.00}}
\put(56.00,8.50){\vector(1,0){3.00}}
\put(75.00,8.50){\vector(1,0){3.00}}
\put(1.50,4.50){\makebox(0,0)[rc]{$A$}}
\put(1.50,12.50){\makebox(0,0)[rc]{$F$}}
\put(14.50,12.50){\makebox(0,0)[lc]{$D$}}
\put(14.50,4.50){\makebox(0,0)[lc]{$C$}}
\put(8.00,1.50){\makebox(0,0)[ct]{$B$}}
\put(8.00,15.50){\makebox(0,0)[cb]{$E$}}
\put(81.50,4.50){\makebox(0,0)[rc]{$A$}}
\put(81.50,12.50){\makebox(0,0)[rc]{$F$}}
\put(94.50,12.50){\makebox(0,0)[lc]{$D$}}
\put(94.50,4.50){\makebox(0,0)[lc]{$C$}}
\put(88.00,1.50){\makebox(0,0)[ct]{$B$}}
\put(88.00,15.50){\makebox(0,0)[cb]{$E$}}
\end{picture}
\end{center}
\caption{}
\label{fig 6-ugol'niki}
\end{figure}
where several ``moves'' of type
$\matrix{
\unitlength=1mm
\special{em:linewidth 0.4pt}
\linethickness{0.4pt}
\begin{picture}(19.00,7.50)
\emline{3.00}{0.00}{1}{0.00}{3.00}{2}
\emline{0.00}{3.00}{3}{4.00}{6.00}{4}
\emline{4.00}{6.00}{5}{7.00}{3.00}{6}
\emline{7.00}{3.00}{7}{3.00}{0.00}{8}
\emline{15.00}{0.00}{9}{12.00}{3.00}{10}
\emline{12.00}{3.00}{11}{16.00}{6.00}{12}
\emline{16.00}{6.00}{13}{19.00}{3.00}{14}
\emline{19.00}{3.00}{15}{15.00}{0.00}{16}
\put(8.00,3.00){\vector(1,0){3.00}}
\emline{3.00}{0.00}{17}{4.00}{6.00}{18}
\emline{12.00}{3.00}{19}{19.00}{3.00}{20}
\end{picture}
}$
are performed one after another. It is not hard to find the
Jacobian for passage from variables
$l_{AC}$, $l_{AE}$, $l_{CE}$ to $l_{BD}$, $l_{BF}$, $l_{DF}$
(where $l_{AC}$ is the length of diagonal $AC$ and so on),
using at each step relations of type~(\ref{eq l3-l6}). We will not write out
the Jacobian, but only the equality between ``state sums''
that follows directly from its form:
\be
\int {\,d \mu(l_{AC}) \cdot \,d \mu(l_{AE}) \cdot \,d \mu(l_{CE}) \over
S_{ABC}\cdot S_{AEF}\cdot S_{CDE}\cdot S_{ACE}} =
\int {\,d \mu(l_{BD}) \cdot \,d \mu(l_{DF}) \cdot \,d \mu(l_{BF}) \over
S_{ABF}\cdot S_{BCD}\cdot S_{DEF}\cdot S_{BDF}},
\label{eq 2-mernye statsummy}
\ee
where one can also multiply the integrands by equal factors of the form
$\exp(\lambda S_{ABCDEF})$. The integration in (\ref{eq 2-mernye statsummy})
can, again, be performed over all flat euclidean hexagon configurations.

In subsection~\ref{subsec 5 tetra} we will compare this simple example to
a similar example for 3-dimensional space.

\section{The case of three and more dimensions}
\label{sec 3-dim}

\subsection{``Elementary transformation'' in 3-dimensional space}
\label{subsec elem-3}

The equality~(\ref{eq l3-l1}) that forms the basis of our transformations
in 2-dimensional space is based on the fact that 6 distances between
four points of a euclidean plane obey one relation.
Similarly, 10 distances between {\em five\/} points of a 3-dimensional
space are bound with one relation (namely, the
Cayley -- Menger equation, see~\cite{B}), and this will allow us to derive
here a ``3-dimensional'' formula analogous to~(\ref{eq l3-l1}).

Let all the distances between points $A,B,C,D,E$ in fig.~\ref{fig ABCDE}
\begin{figure}
\begin{center}
\unitlength=1.00mm
\special{em:linewidth 0.4pt}
\linethickness{0.4pt}
\begin{picture}(120.00,39.00)
\emline{2.00}{18.00}{1}{32.00}{38.00}{2}
\emline{32.00}{38.00}{3}{28.00}{18.00}{4}
\emline{28.00}{18.00}{5}{32.00}{2.00}{6}
\emline{32.00}{2.00}{7}{2.00}{18.00}{8}
\emline{2.00}{18.00}{9}{18.00}{13.00}{10}
\emline{18.00}{13.00}{11}{28.00}{18.00}{12}
\emline{32.00}{2.00}{13}{18.00}{13.00}{14}
\emline{18.00}{13.00}{15}{32.00}{38.00}{16}
\emline{32.00}{38.00}{17}{32.00}{2.00}{18}
\emline{2.00}{18.00}{19}{28.00}{18.00}{20}
\emline{57.00}{38.00}{21}{48.00}{18.00}{22}
\emline{48.00}{18.00}{23}{72.00}{2.00}{24}
\emline{72.00}{2.00}{25}{57.00}{38.00}{26}
\emline{57.00}{38.00}{27}{56.00}{18.00}{28}
\emline{56.00}{18.00}{29}{48.00}{18.00}{30}
\emline{56.00}{18.00}{31}{72.00}{2.00}{32}
\put(47.50,13.50){\vector(2,1){8.00}}
\put(1.00,18.00){\makebox(0,0)[rc]{$A$}}
\put(32.00,39.00){\makebox(0,0)[cb]{$E$}}
\put(27.00,19.00){\makebox(0,0)[rb]{$D$}}
\put(18.00,14.00){\makebox(0,0)[rb]{$C$}}
\put(32.00,1.00){\makebox(0,0)[ct]{$B$}}
\put(55.00,19.00){\makebox(0,0)[rb]{$\alpha$}}
\put(57.00,18.00){\makebox(0,0)[lb]{$\beta$}}
\put(57.00,17.00){\makebox(0,0)[rt]{$\gamma$}}
\put(47.00,18.00){\makebox(0,0)[rc]{$A$}}
\put(57.00,39.00){\makebox(0,0)[cb]{$E$}}
\put(72.00,1.00){\makebox(0,0)[ct]{$B$}}
\put(47.50,13.00){\makebox(0,0)[ct]{$C,D$}}
\put(62.00,30.00){\makebox(0,0)[lb]{\footnotesize
 The same as on the left, projected}}
\put(63.00,26.50){\makebox(0,0)[lb]{\footnotesize
 on a plane orthogonal to $CD$}}
\end{picture}
\end{center}
\caption{}
\label{fig ABCDE}
\end{figure}
be fixed, except $l_{AE}$ and $l_{BE}$. We will derive the dependence
between differentials of those latter with the help of dihedral
angles: $\alpha$ between the faces $ACD$ and $ECD$, and $\beta$
between $ECD$ and $BCD$.
The sum of those angles does not change, because the edge lengths
in tetrahedron $ABCD$ are fixed, thus
\be
\left| d \alpha \right| = \left| d \beta \right|.
\label{eq d alpha d beta}
\ee
To connect $\,d l_{AE}$ with $\,d \alpha$, let us draw heights
$AK$ and $EL$ of faces $ACE$ and $CDE$. It seems convenient to picture
this by {\em unrolling\/}
both faces on the picture plane, see fig.~\ref{fig razvertka}.
\begin{figure}
\begin{center}
\unitlength=1mm
\special{em:linewidth 0.4pt}
\linethickness{0.4pt}
\begin{picture}(45.00,22.00)
\emline{2.00}{8.00}{1}{20.00}{2.00}{2}
\emline{20.00}{2.00}{3}{44.00}{13.00}{4}
\emline{44.00}{13.00}{5}{20.00}{21.00}{6}
\emline{20.00}{21.00}{7}{2.00}{8.00}{8}
\emline{2.00}{8.00}{9}{20.00}{8.00}{10}
\emline{20.00}{2.00}{11}{20.00}{21.00}{12}
\emline{20.00}{13.00}{13}{44.00}{13.00}{14}
\put(1.00,8.00){\makebox(0,0)[rc]{$A$}}
\put(45.00,13.00){\makebox(0,0)[lc]{$E$}}
\put(20.00,22.00){\makebox(0,0)[cb]{$D$}}
\put(20.00,1.00){\makebox(0,0)[ct]{$C$}}
\put(21.00,8.00){\makebox(0,0)[lc]{$K$}}
\put(19.00,13.00){\makebox(0,0)[rc]{$L$}}
\end{picture}
\end{center}
\caption{}
\label{fig razvertka}
\end{figure}
In terms of segment lengths in fig.~\ref{fig razvertka} we get
\be
l_{AE}^2 = l_{AK}^2 + l_{EL}^2 -2l_{AK}l_{EL} \cos\alpha + l_{KL}^2.
\label{eq t-cos}
\ee
As only $l_{AE}$ and $\alpha$ can change,
$$
l_{AE} \,d l_{AE} = l_{AK}\, l_{EL} \sin \alpha \,d \alpha.
$$

Next, the volume of tetrahedron $ACDE$
$$
V_{ACDE}={1\over 6}\, l_{CD}\, l_{AK}\, l_{EL} \sin \alpha,
$$
whence
\be
{l_{AE} \,d l_{AE} \over V_{ACDE}}=6\cdot {\,d\alpha\over l_{CD}}.
\label{eq primechatel'noe}
\ee

Having done similar calculations for tetrahedron $BCDE$ and taking
(\ref{eq d alpha d beta}) into account, we get the desired analog
of formula~(\ref{eq l3-l1}):
\be
\left| l_{AE} \,d l_{AE} \over V_{ACDE} \right| =
\left| l_{BE} \,d l_{BE} \over V_{BCDE} \right|,
\label{eq 12}
\ee
that is
\be
\left| {l_{BE}\over l_{AE}} {\p l_{BE}\over \p l_{AE}}\right| =
\left| V_{BCDE}\over V_{ACDE} \right|.
\label{eq dlBE/dlAE}
\ee

\subsection{Transformations of a complex of five tetrahedra}
\label{subsec 5 tetra}

As an example with more vertices and tetrahedra let us take two
decompositions of a cube $C'BA'DAD'CB'$ in 5 tetrahedra depicted in
fig.~\ref{fig kuby}.
\begin{figure}
\begin{center}
\unitlength=1.00mm
\special{em:linewidth 0.4pt}
\linethickness{0.4pt}
\begin{picture}(101.00,43.00)
\emline{0.00}{10.00}{1}{16.00}{2.00}{2}
\emline{16.00}{2.00}{3}{40.00}{10.00}{4}
\emline{40.00}{10.00}{5}{40.00}{34.00}{6}
\emline{40.00}{34.00}{7}{16.00}{26.00}{8}
\emline{16.00}{26.00}{9}{0.00}{34.00}{10}
\emline{0.00}{34.00}{11}{0.00}{10.00}{12}
\emline{0.00}{34.00}{13}{24.00}{42.00}{14}
\emline{24.00}{42.00}{15}{40.00}{34.00}{16}
\emline{16.00}{26.00}{17}{16.00}{2.00}{18}
\emline{0.00}{10.00}{19}{6.00}{12.00}{20}
\emline{9.00}{13.00}{21}{15.00}{15.00}{22}
\emline{18.00}{16.00}{23}{24.00}{18.00}{24}
\emline{24.00}{18.00}{25}{28.00}{16.00}{26}
\emline{30.00}{15.00}{27}{34.00}{13.00}{28}
\emline{36.00}{12.00}{29}{40.00}{10.00}{30}
\emline{60.00}{10.00}{31}{76.00}{2.00}{32}
\emline{76.00}{2.00}{33}{100.00}{10.00}{34}
\emline{100.00}{10.00}{35}{100.00}{34.00}{36}
\emline{100.00}{34.00}{37}{76.00}{26.00}{38}
\emline{76.00}{26.00}{39}{60.00}{34.00}{40}
\emline{60.00}{34.00}{41}{60.00}{10.00}{42}
\emline{60.00}{34.00}{43}{84.00}{42.00}{44}
\emline{84.00}{42.00}{45}{100.00}{34.00}{46}
\emline{76.00}{26.00}{47}{76.00}{2.00}{48}
\emline{60.00}{10.00}{49}{66.00}{12.00}{50}
\emline{69.00}{13.00}{51}{75.00}{15.00}{52}
\emline{78.00}{16.00}{53}{84.00}{18.00}{54}
\emline{84.00}{18.00}{55}{88.00}{16.00}{56}
\emline{90.00}{15.00}{57}{94.00}{13.00}{58}
\emline{96.00}{12.00}{59}{100.00}{10.00}{60}
\emline{0.00}{34.00}{61}{16.00}{2.00}{62}
\emline{16.00}{2.00}{63}{40.00}{34.00}{64}
\emline{40.00}{34.00}{65}{0.00}{34.00}{66}
\emline{0.00}{34.00}{67}{6.00}{30.00}{68}
\emline{9.00}{28.00}{69}{15.00}{24.00}{70}
\emline{18.00}{22.00}{71}{24.00}{18.00}{72}
\emline{24.00}{18.00}{73}{28.00}{22.00}{74}
\emline{30.00}{24.00}{75}{34.00}{28.00}{76}
\emline{36.00}{30.00}{77}{40.00}{34.00}{78}
\emline{16.00}{2.00}{79}{18.00}{6.00}{80}
\emline{19.00}{8.00}{81}{21.00}{12.00}{82}
\emline{22.00}{14.00}{83}{24.00}{18.00}{84}
\emline{60.00}{10.00}{85}{76.00}{26.00}{86}
\emline{76.00}{26.00}{87}{84.00}{42.00}{88}
\emline{76.00}{26.00}{89}{100.00}{10.00}{90}
\emline{60.00}{10.00}{91}{66.00}{18.00}{92}
\emline{69.00}{22.00}{93}{75.00}{30.00}{94}
\emline{78.00}{34.00}{95}{84.00}{42.00}{96}
\emline{84.00}{42.00}{97}{88.00}{34.00}{98}
\emline{90.00}{30.00}{99}{94.00}{22.00}{100}
\emline{96.00}{18.00}{101}{100.00}{10.00}{102}
\emline{24.00}{42.00}{103}{24.00}{36.00}{104}
\emline{24.00}{33.00}{105}{24.00}{27.00}{106}
\emline{24.00}{24.00}{107}{24.00}{18.00}{108}
\emline{84.00}{42.00}{109}{84.00}{36.00}{110}
\emline{84.00}{33.00}{111}{84.00}{27.00}{112}
\emline{84.00}{24.00}{113}{84.00}{18.00}{114}
\put(-1.00,34.00){\makebox(0,0)[rc]{$A$}}
\put(-1.00,10.00){\makebox(0,0)[rc]{$C'$}}
\put(16.00,1.00){\makebox(0,0)[ct]{$B$}}
\put(41.00,10.00){\makebox(0,0)[lc]{$A'$}}
\put(41.00,34.00){\makebox(0,0)[lc]{$C$}}
\put(24.00,43.00){\makebox(0,0)[cb]{$B'$}}
\put(59.00,34.00){\makebox(0,0)[rc]{$A$}}
\put(59.00,10.00){\makebox(0,0)[rc]{$C'$}}
\put(76.00,1.00){\makebox(0,0)[ct]{$B$}}
\put(101.00,10.00){\makebox(0,0)[lc]{$A'$}}
\put(101.00,34.00){\makebox(0,0)[lc]{$C$}}
\put(84.00,43.00){\makebox(0,0)[cb]{$B'$}}
\put(17.00,27.00){\makebox(0,0)[cb]{$D'$}}
\put(24.00,20.00){\makebox(0,0)[rb]{$D$}}
\put(84.00,17.00){\makebox(0,0)[ct]{$D$}}
\put(79.00,27.00){\makebox(0,0)[lt]{$D'$}}
\emline{60.00}{10.00}{115}{70.00}{10.00}{116}
\emline{75.00}{10.00}{117}{85.00}{10.00}{118}
\emline{90.00}{10.00}{119}{100.00}{10.00}{120}
\end{picture}
\end{center}
\caption{}
\label{fig kuby}
\end{figure}
Surely, this figure has a ``topological'' character --- the true
edge lengths are not reflected in it and, moreover, ``faces'' like $AC'BD'$
are not obliged to be flat, and will be in fact regarded as
tetrahedra with finite volume in intermediate calculations.

Let us transform the LHS of fig.~\ref{fig kuby} in its RHS by doing
moves of the type described in subsection~\ref{subsec elem-3}, i.e.\
replacing one of the edges of diagram with another and calculating
at the same time the partial derivatives using
formula~(\ref{eq dlBE/dlAE}). The product of those derivatives
will be exactly
the Jacobian for passage from 6 cube face diagonal lengths
in the LHS to 6 diagonal lengths in the RHS (the lengths of cube edges
in LHS and RHS are the same, of course).

Our first move will be the following. Note that in the vertex~$A'$
exactly 3 edges intersect, so changing of the length $A'B$ deforms only
one (of five pictured in the LHS of fig.~\ref{fig kuby}) tetrahedron,
namely $A'BCD$. Take now for five points in
subsection~\ref{subsec elem-3} points $A$, $B$, $B'$, $C$ and $D$, and
replace the edge $A'B$ with $A'B'$. After this, the tetrahedron $A'BCD$
will cease to exist, while the tetrahedron $A'B'CD$ will appear (i.e.\
all its 6 edges will be present on the new diagram). We will not
write down the corresponding formulae of type~(\ref{eq dlBE/dlAE}),
but will write down in full the 12 transformations
needed for passing from LHS of fig.~\ref{fig kuby} to its RHS.
Below in the LHS we indicate which edge is replaced with which one,
and in the RHS --- which tetrahedra disappear and appear:
\begin{eqnarray*}
A'B  \to A'B' \,, && A'BCD    \to A'B'CD \;;  \\
BC'  \to B'C' \,, && ABC'D    \to AB'C'D \;;  \\
BD'  \to B'D' \,, && ABCD'    \to AB'CD' \;;  \\[\smallskipamount]
BC   \to BC' \,,  && ABCD     \to ABC'D \;;   \\
CD'  \to C'D'\,,  && AB'CD'   \to AB'C'D' \;; \\
A'C  \to A'C'\,,  && A'B'CD   \to A'B'C'D \;; \\[\smallskipamount]
CD   \to CD' \,,  && AB'CD    \to AB'CD' \;;  \\
A'D  \to A'D'\,,  && A'B'C'D  \to A'B'C'D' \;;\\
BD   \to BD' \,,  && ABC'D    \to ABC'D' \;;  \\[\smallskipamount]
AD   \to A'D \,,  && AB'C'D   \to A'B'C'D \;; \\
AB   \to A'B \,,  && ABC'D'   \to A'BC'D' \;; \\
AC   \to A'C \,,  && AB'CD'   \to A'B'CD' \;.
\end{eqnarray*}

Note that after the first three moves there remain only 3 edges out of
6 intersecting originally in the vertex~$B$ --- but, in return,
3 new ones arise at the vertex~$B'$. Then the pairs $C$ and $C'$,
$D$ and $D'$, $A$ and $A'$ undergo similar changes.

One can check that finally the following formula, an analog
of (\ref{eq 2-mernye statsummy}), comes out:
\be
\int \prod_{\hbox{\scriptsize$\matrix{
\hbox{over 5 initial}\cr
\hbox{tetrahedra} }$}} V^{-1}
\prod_{\hbox{\scriptsize$\matrix{
\hbox{over 6 initial}\cr
\hbox{diagonals} }$}} d\mu(l) =
\int \prod_{\hbox{\scriptsize$\matrix{
\hbox{over 5 final}\cr
\hbox{tetrahedra} }$}} V^{-1}
\prod_{\hbox{\scriptsize$\matrix{
\hbox{over 6 final}\cr
\hbox{diagonals} }$}} d\mu(l).
\label{eq 3-mernye statsummy}
\ee

Again, in analogy with section~\ref{sec 2-dim}, one can integrate
in both sides over all possible configurations of euclidean
eight-vertex figure
with fixed lengths of ``nondiagonal'' edges, assuming that all
$V$ and $\,d\mu$ are always positive.

\subsection{An example with ``curvature''}
\label{subsec krivizna}

Return now to constructions of subsection~\ref{subsec elem-3} in order
to show that one can introduce there a ``discrete curvature concentrated
in the edges of the complex'', in the spirit of Regge
calculus~\cite{Regge calculus} in quantum gravity. Namely, let
the ``curvature'' in fig.~\ref{fig ABCDE}
be concentrated in edge $CD$, which shows itself in the fact that the sum
of dihedral angles $\alpha$ between faces $ACD$ and $ECD$, $\beta$
between $ECD$ and $BCD$, and $\gamma$ between $BCD$ and $ACD$ equals not
$2\pi$ but another fixed value, $2\pi-\omega$. The Cayley -- Menger
equation for distances between vertices in fig.~\ref{fig ABCDE}
is replaced by another relation depending on $\omega$. This relation
can be derived by adding to (\ref{eq t-cos}) two similar equalities
expressing $l_{BE}$ through $\beta$ and $l_{AB}$ through $\gamma$,
computing the triangle heights (two of the three heights are pictured
in ``unrolled'' form in fig.~\ref{fig razvertka})
and segments in which they divide the triangles' base $CD$ in terms of
side lengths using usual planimetry, and then excluding $\alpha$, $\beta$
and $\gamma$ that obey
$$
\alpha + \beta + \gamma = 2\pi - \omega
$$
from the obtained relations.
Important is that formula $\left|d \alpha\right|=\left| d \beta
\right|$ (\ref{eq d alpha d beta}) and all subsequent calculations
in subsection~\ref{subsec elem-3}, including formula (\ref{eq dlBE/dlAE}),
remain valid, with the volumes of tetrahedra expressed through
edge lengths by the usual formula:
\begin{eqnarray*}
V^2={1\over 144}\bigl(a_1a_5(a_2+a_3+a_4+a_6-a_1-a_5)+
a_2a_6(a_1+a_3+a_4+a_5-a_2-a_6)\\
+a_3a_4(a_1+a_2+a_5+a_6-a_3-a_4)- a_1a_2a_4-a_2a_3a_5-a_1a_3a_6-a_4a_5a_6
\bigr),
\end{eqnarray*}
where if e.g. $V=V_{ACDE}$ is the volume of tetrahedron $ACDE$, then
$$
a_1=l_{AE}^2,\quad a_2=l_{CE}^2,\quad a_3=l_{DE}^2,\quad
a_4=l_{AC}^2,\quad a_5=l_{CD}^2,\quad a_6=l_{AD}^2,
$$
but the ten lengths enjoy now the relation including the curvature!

\subsection{Multidimensional generalizations}

In full analogy with the ``3-dimensional'' formula
(\ref{eq primechatel'noe}),
one can obtain a similar formula for $n$-dimensional space:
\be
{l_{AZ}\,dl_{AZ}\over V_{ACD\ldots Z}}=n(n-1)\,{d\alpha\over
S_{\overline{AZ}}},
\label{eq 2.4 ugol}
\ee
where $V_{ACD\ldots Z}$ is now the $n$-dimensional hypervolume of the
corresponding simplex (with $n+1$ vertices), $l_{AZ}$ is the length of
edge $AZ$, $S_{\overline{AZ}}$ is the
$(n-2)$-dimensional ``hyperarea'' of the simplex obtained from
$ACD\ldots Z$ by moving off the vertices $A$ and $Z$, and $\alpha$ is
the angle between two $(n-1)$-dimensional faces
obtained from $ACD\ldots Z$ by moving off points
$A$ and $Z$ respectively. From here, of course, analogs of formulae
(\ref{eq 12}) and (\ref{eq dlBE/dlAE}) follow, e.g., the analog
for the first of them is
\be
\left|l_{AZ}\,dl_{AZ}\over V_{ACD\ldots Z}\right|=
\left|l_{BZ}\,dl_{BZ}\over V_{BCD\ldots Z}\right|.
\label{eq 2.4 dliny}
\ee

\section{Discussion}
\label{questions}

We have presented in subsection~\ref{subsec 5 tetra} only one of
the simplest examples of reconstruction of a 3-dimensional
simplicial complex, and in subsection~\ref{subsec krivizna} ---
only a first indication that the ``discrete curvature'' understood
in the sense of Regge calculus in quantum gravity can be included
in our models. Thus the first natural problems for further work are:
describe reconstructions leading to state sum equalities
of type (\ref{eq 3-mernye statsummy}) for more complicated complexes,
preferably including edges where nonzero curvature is concentrated.
Maybe one should consider formulae like
(\ref{eq primechatel'noe}) and (\ref{eq 2.4 ugol}) as defining one more
``duality'' --- between an edge length and an angle.

Another group
of questions is about whether there exist ``moves'' of some other form
than presented in this paper that conserve a properly chosen
``state sum'', say, in two dimensions --- moves of type
$\matrix{
\unitlength=1.00mm
\special{em:linewidth 0.4pt}
\linethickness{0.4pt}
\begin{picture}(17.00,7.00)
\emline{0.00}{0.00}{1}{3.00}{6.00}{2}
\emline{3.00}{6.00}{3}{6.00}{0.00}{4}
\emline{6.00}{0.00}{5}{0.00}{0.00}{6}
\emline{11.00}{0.00}{7}{14.00}{6.00}{8}
\emline{14.00}{6.00}{9}{17.00}{0.00}{10}
\emline{17.00}{0.00}{11}{11.00}{0.00}{12}
\emline{11.00}{0.00}{13}{14.00}{2.00}{14}
\emline{14.00}{2.00}{15}{14.00}{6.00}{16}
\emline{14.00}{2.00}{17}{17.00}{0.00}{18}
\put(9.00,3.00){\vector(-1,0){2.50}}
\put(8.00,3.00){\vector(1,0){2.50}}
\end{picture}
}$
and so on.

The third group
of questions is: in this paper, {\em one\/} of two 2-dimensional models
from~\cite{KS} is generalized. It is interesting to try and find
a multidimensional generalization of the second model (related to
Veneziano amplitude) and, more generally, for wider variety of models.

The fourth group
of questions, very natural for those working in the theory of
exactly solvable models, deals with actual calculation of ``state
sums''. Note that the paper~\cite{korepanov k-tetraedr} indicates,
in particular, at the closeness of the 2-dimensional duality and
Yang-Baxter equations.

The fifth group
of questions is about models that can be obtained as reductions
of models in a higher-dimensional space. As soon as the dimensionality
of a space can grow up to infinity, it seems likely that the richness
of ``low-dimensional'' models can, too, increase considerably. Contiguous
is the
question about including in multidimensional amplitudes
an exponential factor like the one in~(\ref{ampl s exp}).

\end{document}